\documentclass[10pt,letterpaper]{article}
\usepackage{graphics}% Include figure files
\usepackage{graphicx}% Include figure files
\usepackage{dcolumn}% Align table columns on decimal point
\usepackage{bm}% bold math

\newcommand{\de}{\partial}

\newcommand{\eq}[2]{\begin{equation} \label{#1} #2 \end{equation}}
\newcommand{\eps}{\epsilon}

\newcommand{\etal}{{\em et al.}}

\newcommand{\diverg}[1]{\nabla\cdot #1}
\newcommand{\rot}[1]{\nabla\times #1}

\newcommand{\EE}{\mathbf{E}}
\newcommand{\eee}{\mathbf{e}}
\newcommand{\fff}{\mathbf{f}}
\newcommand{\hhh}{\mathbf{h}}

\newcommand{\rr}{\mathbf{r}}
\newcommand{\rrort}{\mathbf{r}_{\bot}}

\newcommand{\DD}{\mathbf{D}}
\newcommand{\HH}{\mathbf{H}}
\newcommand{\PP}{\mathbf{P}}
\newcommand{\SSS}{\mathbf{S}}

\newcommand{\PNL}{\mathbf{P}_{\rm NL}}

\begin{document}

%% NOTE: TITLE PAGE & TOC NOT USED FOR MANUSCRIPT SUBMISSIONS %%
%\title{Template and style guide for authors submitting to \textit{Optics Express}}

%\vskip4pc

%\tableofcontents
%\clearpage
%% NO TITLE PAGE FOR OPEX SUBMISSIONS %%

%% START HERE
%%%%%%%%%%%%%%%%%% title page information %%%%%%%%%%%%%%%%%%
\title{\bf An accurate envelope equation for light propagation in
photonic nanowires: new nonlinear effects}

\author{Truong X. Tran and Fabio Biancalana \\ Max Planck Institute for the Science of Light, \\
G\"{u}nther-Scharowsky-Str. 1/Bau 24, 91058 Erlangen, Germany}

\maketitle

\begin{abstract} We derive a set of new unidirectional evolution equations for
photonic nanowires, i.e. waveguides with sub-wavelength core diameter. Contrary to previous approaches,
our formulation simultaneously takes into account both the vector
nature of the electromagnetic field and the full variations of the
effective modal profiles with wavelength.
This leads to the discovery of new, previously unexplored nonlinear
effects which have the potential to affect soliton propagation
considerably.
We specialize our theoretical considerations to the case of perfectly
circular silica strands in air, and we support our analysis with
detailed numerical simulations. \end{abstract}

% REPLACE WITH CORRECT OCIS CODES FOR YOUR ARTICLE

%%%%%%%%%%%%%%%%%%%%%%%%%%  body  %%%%%%%%%%%%%%%%%%%%%%%%%%
\section{Introduction}
\label{introduction}
Evolution equations that are able to accurately describe the
propagation of ultra-short pulses in strongly nonlinear media are
extremely valuable tools in modern computational nonlinear optics, and
especially in fiber optics \cite{agrawalbook}.
On one hand, it is always possible to directly solve Maxwell's
equations in presence of linear dispersion and nonlinear polarization
\cite{allentaflovebook}, capturing the full complexity of the
nonlinear propagation, but in the majority of cases
very large computational resources are needed for this, which means
that one can only effectively simulate short propagation distances
along an optical fiber. On the other hand, it is possible to
drastically reduce the complexity of Maxwell's equations by using the
envelopes of the electromagnetic (EM) fields, while still maintaining
an accurate description of the important dynamical degrees of freedom.
This can lead to different approximation schemes based on various
different versions of the nonlinear Schr\"odinger equation (NLSE),
that have been used quite successfully in the past to describe a
variety of linear and nonlinear effects in large-core optical fibers.
Amongst them, we can mention phenomena such as soliton propagation
\cite{hasegawakodama}, modulational instabilities and four-wave mixing
\cite{modulationalinstability}, and third and higher-order dispersive
effects \cite{agrawalbook}. A specific extended version of the
NLSE (also called the generalized NLSE, or GNLSE for brevity), which
also includes terms describing the Raman effect \cite{gordon},
self-steepening \cite{agrawalbook} and the full complexity of the
group velocity dispersion (GVD) of the waveguide, has been used in
recent times to describe the important phenomenon of supercontinuum
generation (SCG),  i.e. the generation of a broad coherent spectrum from a short input pulse,
with spectacular success \cite{gaeta}. This equation
has led to a number of important advances in the theoretical
understanding of SCG in ultra-small core fibers such as tapered fibers
(TFs) \cite{taperedfibers} and photonic crystal fibers (PCFs)
\cite{russell,dudley}. In particular the phenomena of emission of
solitonic dispersive radiation (also called resonant radiation or
Cherenkov radiation \cite{akhmediev}), and the stabilization of the
Raman self-frequency shift (RSFS) of solitons by means of such
radiation \cite{scienceskryabin,biancalana}, are recent important
advances that have been possible only by means of a thorough
theoretical understanding of the structure of GNLSE.

Equations like the GNLSE are based on several assumptions and
approximations. The most important {\em approximation} used in all
NLSE-based evolution equations is the slowly-varying envelope
approximation (SVEA) in both space and time \cite{brabec}. This
amounts to decoupling the fast oscillating spatial and temporal
frequencies from the envelope function, which is the only complex
function used to describe the dynamics. This reduction is only
possible if the envelope profile contains many oscillations, which is
typically the case in the visible and mid infrared regime when the
pulse duration is larger than a few tens of femtoseconds. SVEA allows
to reduce the second-order wave equation for the electric field into a
first-order equation (which only contains the forward-propagating
modes and neglects the backward-propagating ones) in which the
evolution coordinate is $z$, the longitudinal spatial coordinate along
the fiber. In this way, the computational complexity of numerical
simulations is drastically reduced.
The most crucial {\em assumption} on which the GNLSE is based is that
the $z$-component of the electric field is very small with respect to
its transverse components (weak guidance regime). This allows to
approximately decouple the transverse dynamics from the longitudinal
dynamics. The weak guidance regime represents a very good assumption
for large-core fibers (when the core size is much larger than the
wavelength of light) and for fibers with a low contrast between the
refractive indices of the core and the cladding.

Recently, however, new types of waveguides with a sub-wavelength size
of the core (broadly referred as {\em photonic nanowires}) have become
accessible to fabrication, which both possess a complex cladding
structure, that allows one to support solid cores with a
sub-wavelength diameter, and have a strong contrast between the
refractive indices of the core and the cladding. To this class of
waveguides belong some specific examples of PCFs, such as extruded
PCFs \cite{mercedes}; the extremely small interstitial features of Kagome
hollow-core PCFs \cite{benabid}; and TFs, i.e. silica rods with
sub-micron diameters surrounded by air or vacuum \cite{taperedfibers}.
As mentioned, the common feature of all these structures is the
non-negligible nature of the longitudinal component of the electric
field (strong guidance regime) \cite{monro1}. The importance of such
component in the dynamics of light has not been clearly recognized in
the past, until very recently \cite{monro1,monro2}. In these latter
works the authors demonstrate that the calculation of the nonlinear
coefficient of photonic nanowires performed in the scalar theory
\cite{foster1,foster2,zheltikov} underestimates its real magnitude of
approximately a factor $2$, while the correct result can be found by
using the more complete vector theory \cite{monro1}. Such interesting
conclusion is also supported by recent experiments performed on
ultra-small core, highly nonlinear tellurite PCFs
\cite{conferencemonro}.

% motivations statement of the paper (net effect is that nonlinearity is reduced)
Inspired by these recent works, in the present paper we wish to
further advance the theoretical understanding of the evolution
equations in sub-wavelength core structures. Here we derive a new,
logically self-consistent forward-evolution equation for photonic
nanowires, based on SVEA, that {\em simultaneously} takes into account
(i) the vector nature of the EM field (polarization and $z$-component
included), (ii) the strong dispersion of the nonlinearity inside the
spectral body of the pulse (i.e. the variation of the effective modal
area as a function of frequency), and (iii) the full variations of the mode
profiles with frequency. In all relevant previous works on the subject
\cite{brabec,monro1,monro2,mamyshev,kolesik,laegsgaard}, all these
properties were not included in the derivation at the same time.
Considering (i), (ii) and (iii) altogether and with a minimal amount of assumptions, leads to new qualitative
results that have not been possible to investigate with previous
formulations. In particular, we demonstrate through a series of
accurate simulations that (iii) leads to the emergence of {\em new nonlinear
terms} that are also indirectly responsible for appreciable modifications
in the dynamics of RSFS, in particular a suppression of the latter. Thus, even though the vector nature of the EM
field in photonic nanowires may lead to an enhancement of the
nonlinear coefficient in nanowires \cite{monro1}, our results show
that this is actually counterbalanced by terms and effects that
partially suppress such enhancement. However, our findings are not in
contradiction with previous works on the subject, for reasons that we
shall explain in the text.

The structure of the paper is the following. In section
\ref{masterequation} we detail the derivation of our master equation,
Eq. (\ref{finalx1}). In section \ref{symmetries} we describe the
symmetries of the master equation, and we demonstrate that our model
coincides with previously published models under various limiting
cases. In section \ref{neweffects} we approximate Eq. (\ref{finalx1})
for long pulses, and find that new nonlinear terms arise, which were
not considered in any previously published formulation. Such terms
become increasingly important when decreasing the size of the core,
and for a given core diameter their impact is maximized around a certain optimal wavelength.
Finally, in section \ref{numericalanalysis} we show results of
numerical simulations, which underline the difference between our new
theory and other known models. Conclusions are given in section
\ref{conclusions}.

% derivation of master equation
\section{Derivation of master equation}
\label{masterequation}
In this section we present a detailed derivation of our master
equation, Eq. (\ref{finalx1}).
Our starting point are Maxwell's equations written for a non-magnetic
medium in the Lorentz-Heaviside units system \cite{jacksonbook}:
\begin{eqnarray}
\rot\HH=\frac{1}{c}\de_{t}\DD+\frac{1}{c}\de_{t}\PNL \label{max1}\\
\rot\EE=-\frac{1}{c}\de_{t}\HH \label{max2}\\
\diverg\HH=0,\ \ \diverg\DD=0, \label{max3}
\end{eqnarray}
where $\EE(\rr,t)$ and $\HH(\rr,t)$ are the electric and magnetic
field respectively, $t$ is time, $\rr$ is the coordinate vector, $c$
is the speed of light in vacuum and $\PNL(\rr,t)$ is the nonlinear
polarization field that we shall specify later.
The displacement field is given by $\DD(\rr,t)=\eps\otimes\EE$, where
$\eps(\rrort,t)$ is the dielectric function that models the linear
response of the medium, i.e. its dispersion. Its Fourier transform
$\eps(\rrort,\omega)$ gives the square of the refractive index along
the transverse direction $\rrort$. The symbol $\otimes$ indicates a
convolution: $a\otimes
b\equiv\int_{-\infty}^{+\infty}a(t-t')b(t')dt'=\int_{-\infty}^{+\infty}b(t-t')a(t')dt'\equiv
b\otimes a$. In the frequency domain, one has
$\DD(\rr,\omega)=\eps(\rrort,\omega)\EE(\rr,\omega)$.

\begin{figure}[htb]
\centering\includegraphics[width=7cm]{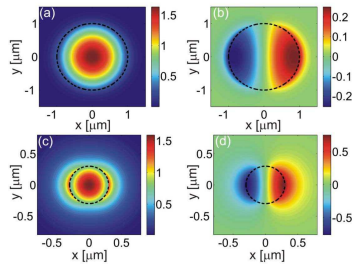}
\caption{\small (Color online) (a,b) Respectively transverse and
longitudinal components of electric field in a silica strand in air of
diameter $d=2$ $\mu$m, for a pump wavelength $\lambda=1$ $\mu$m. (c,d)
The same as (a,b), but for diameter $d=0.6$ $\mu$m. The silica core is
indicated with a dashed black line. Note the expulsion of the total field from the core towards the
low-index medium. Also, the magnitude of the longitudinal component in
case (c,d) is more than 30\% of the transverse component.}  \label{fig1}
\end{figure}

In absence of nonlinear polarization ($\PNL=0$), the fiber has linear
eigenmodes in the form of plane waves, solutions of the linearized
Maxwell equations at a specified frequency $\omega$ \cite{snyderlove}.
These are given by:
\eq{eigen1}{\EE_{m,\omega}(\rr,t)=\frac{\eee_{m,\omega}(\rrort)}{\sqrt{N_{m\omega}}}e^{i\beta_{m}(\omega)z-i\omega
t}}
and \eq{eigen2}{\HH_{m,\omega}(\rr,t)=\frac{\hhh_{m,\omega}(\rrort)}{\sqrt{N_{m\omega}}}e^{i\beta_{m}(\omega)z-i\omega
t},} where $m$ is a collective index that indicates both the
polarization state and the mode number, and $\beta_{m}(\omega)$ is the
propagation constant. $N_{m\omega}$ is a mode- and frequency-dependent normalization constant
that will be defined shortly.
Functions $\eee_{m,\omega}(\rrort)$ and $\hhh_{m,\omega}(\rrort)$ give the
profile of the fiber eigenmodes
\cite{snyderlove}. Note that $\eee, \hhh$ have all three components,
including a $z$-component which becomes comparable with the other two
transverse components for a sufficiently small core and for a
relatively large contrast between the core and the cladding refractive
indices, as pointed out in Ref. \cite{monro1}. Fig. \ref{fig1} shows
contour plots of the transverse and the longitudinal components of the electric field
for two circular silica strands of diameters $d=2$ $\mu$m [Fig.
\ref{fig1}(a,b)] and $d=0.6$ $\mu$m [Fig. \ref{fig1}(c,d)]
respectively, for a pump wavelength $\lambda=1$ $\mu$m. Note the
gradual expulsion of the total (transverse and longitudinal) field
from the core towards the low-index medium (air in this case). For
PCFs and for fibers with a highly non-trivial cladding structure, the
vector profiles of the electric and magnetic eigenmodes must be
calculated numerically by means of a mode solver, but for the specific
case of optical fibers with a perfectly circular core, $\eee$ and
$\hhh$ are known analytically, provided that the propagation constant
is known, see Ref. \cite{snyderlove}.
\begin{figure}[htb]
\centering\includegraphics[width=7cm]{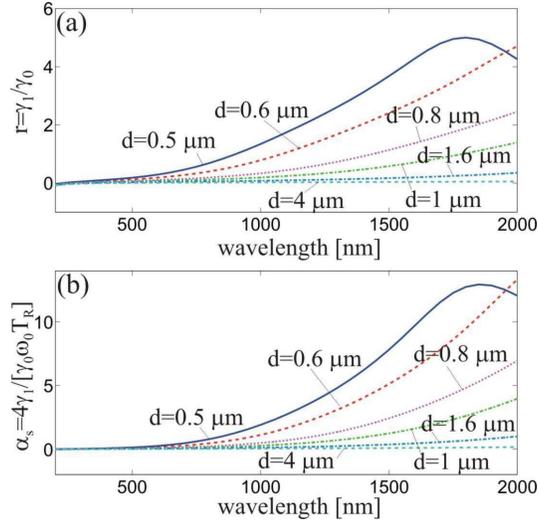}
\caption{\small (Color online) (a) Plot of the dimensionless
coefficient $r\equiv\gamma_{1}/\gamma_{0}=\Gamma^{1000}/\Gamma^{0000}$ as a
function of wavelength, for silica strands in air of diameters
$d=0.5, 0.6, 0.8, 1, 1.6$ and $4$ $\mu$m in the fundamental mode of
one polarization ($m=1$). This function
is predominantly positive in a broad range of wavelengths. $r$ quantifies the relative
importance between the new nonlinear term T3 in Eq. (\ref{approx1})
and the Kerr effect term T1. (b) Plot of the dimensionless coefficient
$\alpha_{S}\equiv
4\Gamma^{1000}/[\Gamma^{0000}\omega_{0}T_{R}]=4\gamma_{1}/[\gamma_{0}\omega_{0}T_{R}]$
as a function of wavelength. $\alpha_{S}$ quantifies the relative
importance between the new nonlinear term T3 in Eq. (\ref{approx1})
and the Raman effect term T2. Both $r$ and $\alpha_{S}$ tend to zero at all wavelengths when progressively
increasing the core diameter.} \label{fig2}
\end{figure}

The full fields are expanded in terms of the eigenmodes as follows:
\begin{eqnarray}
\EE(\rr,t)=\frac{1}{2\pi}\sum_{m}\int d\omega
A_{m\omega}(z)\frac{\eee_{m,\omega}(\rrort)}{\sqrt{N_{m\omega}}}e^{i\beta_{m}(\omega)z-i\omega t}
\label {exp1}\\
\HH(\rr,t)=\frac{1}{2\pi}\sum_{m}\int d\omega
A_{m\omega}(z)\frac{\hhh_{m,\omega}(\rrort)}{\sqrt{N_{m\omega}}}e^{i\beta_{m}(\omega)z-i\omega
t}\label {exp2},
\end{eqnarray}
where we have used the convention $f(t)=[2\pi]^{-1}\int d\omega
f(\omega)e^{-i\omega t}$ for the direct Fourier transform and
$f(\omega)=\int dt f(t)e^{i\omega t}$ for its inverse. It is customary to define the normalized field $\hat{\eee}_{m\omega}\equiv\eee_{m\omega}/\sqrt{N_{m\omega}}$ (and similarly for the
magnetic field), in order to have an orthonormal basis on which to expand any physical field, analogously to what is done in quantum mechanics.
We further note
that in Ref. \cite{monro1} the fields are expanded by assuming that
the modes $\hat{\eee}_{m\omega}$ and $\hat{\hhh}_{m\omega}$ are unchanged around a
reference frequency $\omega_{0}$, which is arbitrary and taken to be
equal to the central pulse frequency. Thus in Ref. \cite{monro1} it is assumed that all spectral components inside
the pulse will have the same mode profile, and there
is no integration over the frequencies analogous to Eqs.
(\ref{exp1}-\ref{exp2}), which is instead considered in Refs.
\cite{kolesik,laegsgaard,mamyshev}. In the present work we wish to be able to describe
the full variations of the mode profiles with frequency because, as it was already demonstrated in
Ref. \cite{dudleyarea}, this actually leads to a large dispersion of the
nonlinearity that strongly affects the dynamics of the resulting
equations, and, most importantly, new nonlinear effects that we shall analyze in detain in section \ref{neweffects}.

The correct orthogonality relations between the linear eigenmodes are
given by the following cross product integral
\cite{snyderlove}:
\eq{orto1}{\frac{1}{2}\int\left[\eee_{j\omega}(\rr_{\perp})\times\hhh^{*}_{k\omega}(\rr_{\perp})\right]\cdot\hat{z}d\rr_{\perp}=\frac{1}{2}\int\left[\eee^{*}_{k\omega}(\rr_{\perp})\times\hhh_{j\omega}(\rr_{\perp})\right]\cdot\hat{z}d\rr_{\perp}=\delta_{jk}N_{j\omega}}
where $N_{j\omega}$ is a normalization constant. However, what is ultimately important is not the absolute normalization,
but the relative change of the normalization constant (and thus of the
modal effective area) with wavelength. That is why we decide to use in the following, without loss of generality, the
orthonormal quantities $\hat{\eee}$, $\hat{\hhh}$, which satisfy the simpler conditions
\eq{orto2}{\frac{1}{2}\int\left[\hat{\eee}_{j\omega}(\rr_{\perp})\times\hat{\hhh}^{*}_{k\omega}(\rr_{\perp})\right]\cdot\hat{z}d\rr_{\perp}=\delta_{jk}.}

It is customary \cite{monro1,kolesik} to define a generalized Poynting
vector as follows:
\eq{poynt1}{\SSS_{m\omega}(\rr,t)=\frac{1}{2}\eps_{0}c\left[
\EE_{m\omega}(\rr,t)\times\HH^{*}(\rr,t)+\EE^{*}(\rr,t)\times\HH_{m\omega}(\rr,t)
\right].}
By using Maxwell's equations (\ref{max1}-\ref{max3}) and the vector
theorem $\diverg(\mathbf{a}\times\mathbf{b})=\mathbf{b}\cdot(\rot{\mathbf{a}})-\mathbf{a}\cdot(\rot{\mathbf{b}})$,
one can find the divergence of $\SSS$:
\eq{poynt2}{\diverg\SSS_{m\omega}=\frac{1}{2}\eps_{0}\left\{
i\omega\HH^{*}\cdot\HH_{m\omega}-\EE_{m\omega}\cdot\de_{t}\DD^{*}-\EE_{m\omega}\cdot\de_{t}\PP^{*}_{\rm
NL}
-\HH_{m\omega}\cdot\de_{t}\HH^{*}+i\omega\eps_{\omega}\EE^{*}\cdot\EE_{m\omega}
\right\}}
After integrating both members of Eq. (\ref{poynt2}) in the transverse
dimensions, performing a time integration as in \cite{kolesik}, and by
 using the reciprocity theorem \cite{snyderlove} one obtains the
following forward-propagating equation (first derived by Kolesik
\etal\  in 2004, see Refs. \cite{kolesik}):
\eq{prop1}{\de_{z}A_{m\omega}=\frac{i\omega}{4c}e^{-i\beta_{m\omega}z}\int\hat{\eee}_{m\omega}^{*}\cdot\PP_{{\rm
NL},\omega}(\rr)d\rr_{\perp}.}
Equation (\ref{prop1}) is the basic equation for the common Fourier
component of the EM fields, on which any forward-propagating theory
must be based. Note that Eq. (\ref{prop1}) is written for a generic
frequency
$\omega$, thus no arbitrary reference frequency has been selected up
to this point. Attempts to directly solve Eq. (\ref{prop1}) have been
given, in various forms, in Refs. \cite{kolesik} and
\cite{laegsgaard}. However, this approach, although exact in absence of backward-propagating modes in the waveguide, does not
provide a transparent understanding of the physical processes
involved, and it does not lead to any serious reduction of the
computational costs with respect to solving directly Maxwell's
equations.

To further progress from Eq. (\ref{prop1}), one needs to write
explicitly the nonlinear polarization field for silica in terms of the
electric field components. This can be done in several ways, depending
on which particular medium or model one is interested in. For
instance, in Ref. \cite{monro1} the authors consider the vector Kerr
nonlinearity of silica in absence of Raman effect. In Ref.
\cite{monro2} the Raman effect has been included, but emphasis was
given to the analysis of stimulated Raman scattering. For our purposes
we choose the following form for the nonlinear polarization of silica
(see also Ref. \cite{laegsgaard}):
$$\PP_{\rm NL}=\frac{1}{4}\chi^{(3)}_{xxxx}(\rrort)\Big{\{} \EE(t)\int
R(t-t_{1})[\EE(t_{1})\cdot\EE^{*}(t_{1})]dt_{1}+\int
R(t-t_{1})\EE(t_{1})[\EE(t)\cdot\EE^{*}(t_{1})]dt_{1} +$$
\eq{pnl1}{+\int R(t-t_{1})\EE^{*}(t_{1})[\EE(t)\cdot\EE(t_{1})]dt_{1} \Big{\}},}
where $\chi^{(3)}_{xxxx}(\rrort)$ is the third-order susceptibility
coefficient, which is the only approximately independent component of
the susceptibility tensor that survives the symmetries
\cite{boydbook}, and is a function of the transverse coordinates.
$R(t)$ is the Raman-Kerr convolution function that describes both the
instantaneous (Kerr) and non-instantaneous (Raman) nonlinearities in
the medium. In the case of silica one has:
\eq{ramankerr1}{R(t)=(1-\theta)\delta(t)+\theta\Theta(t)h(t),} where
$h(t)\equiv[\tau^{2}_{1}+\tau^{2}_{2}][\tau_{1}\tau_{2}^{2}]^{-1}\exp(-t/\tau_{2})\sin(t/\tau_{1})$
is the Raman response function set by the vibrations of silica
molecules induced by the EM field, with $\tau_{1}=12.2$ fs and
$\tau_{2}=32$ fs. Coefficient $\theta$ parameterizes the relative
importance between Raman and Kerr effect in Eq. (\ref{ramankerr1}),
and the experimental value for silica is $\theta\simeq0.18$
\cite{agrawalbook}. In Eq. (\ref{ramankerr1}), $\delta(t)$ is the
Dirac function and $\Theta(t)$ is a Heaviside (step) function that is
introduced to preserve causality. In addition, the response function
is normalized in such a way that $\int_{-\infty}^{+\infty}R(t)dt=1$.
Other, more precise forms of the Raman response functions for silica
are known \cite{agrawalbook}. In a more accurate formulation, the
isotropic and the anisotropic parts of the nuclear response should be
included \cite{hellwarth}. However, the following discussions are
largely independent of the precise form of $R(t)$, and are valid in
general for any functional form of the response function.

After substituting expression (\ref{pnl1}) into Eq. (\ref{prop1}), we
explicitly obtain:
\begin{eqnarray}
&&\de_{z}A_{m\omega}=\frac{i\omega}{4c}\frac{1}{4}\frac{1}{(2\pi)^{2}}\int
\chi^{(3)}_{xxxx} d\rr_{\perp}\Big{(}\sum_{kdl}\int d\omega' d\omega''
R(\omega-\omega')A_{k,\omega'}A_{l,\omega''}A^{*}_{d,\omega'+\omega''-\omega}\cdot \nonumber\\
&&e^{i(\beta_{k,\omega'}+\beta_{l,\omega''}-\beta_{d,\omega'+\omega''-\omega}-\beta_{m\omega})z}
\Big{[}(\hat{\eee}^{*}_{m\omega}\cdot\hat{\eee}_{k\omega'})(\hat{\eee}_{l\omega''}\cdot\hat{\eee}^{*}_{d\omega'+\omega''-\omega})
+(\hat{\eee}^{*}_{m\omega}\cdot\hat{\eee}_{l\omega''})(\hat{\eee}_{k\omega'}\cdot\hat{\eee}^{*}_{d\omega'+\omega''-\omega})+ \nonumber\\
&&(\hat{\eee}^{*}_{m\omega}\cdot\hat{\eee}^{*}_{d,\omega'+\omega''-\omega})(\hat{\eee}_{k\omega'}\cdot\hat{\eee}_{l\omega''})
\Big{]}\Big{)}.\label{substx1}
\end{eqnarray}
Only at this point we choose an arbitrary reference frequency
$\omega_{0}$; furthermore, in order to capture the variations in the
mode profile around $\omega_{0}$, we expand the linear modes $\hat{\eee}$ in
Taylor series:
\eq{expansion1}{\hat{\eee}_{m\omega}=\sum_{j\geq
0}\frac{1}{j!}\fff^{(j)}_{m,\omega_{0}}(\rr_{\perp})\left(\frac{\omega-\omega_{0}}{\omega_{0}}
\right)^{j}=\sum_{j\geq
0}\frac{1}{j!}\left[\omega_{0}^{j}\frac{\de^{j}\hat{\eee}_{m,\omega}(\rr_{\perp})}{\de\omega^{j}}\right]_{\omega=\omega_{0}}\left(\frac{\Delta\omega}{\omega_{0}}
\right)^{j},} where $\Delta\omega\equiv\omega-\omega_{0}$ is the
frequency detuning from the reference frequency $\omega_{0}$, and the quantity $\fff^{(j)}_{m,\omega_{0}}$ is given by the
term inside the square brackets in the last member of Eq.
(\ref{expansion1}), and is proportional to the $j$-th derivative of
the mode profile.
A Taylor expansion equivalent to our Eq. (\ref{expansion1}) was also
proposed in Eqs. (28) and (37) of Ref. \cite{laegsgaard}. However, in
this latter work the author (being mostly interested in large core,
low-contrast PCFs) gets rid of the longitudinal components of the EM
fields [see his Eq. (13)], analyzes the scalar case, and does not
elaborate further on the physical meaning of the corrections induced
by the higher-order terms. Expansion (\ref{expansion1}) is also not considered in Refs. \cite{monro1,poletti}.
In our derivation, the expansion
(\ref{expansion1}) is a crucial step. In fact, by taking into account
the modifications of the quasi-linear mode profiles, we are led to
a correct identification of new nonlinear effects
that, under certain circumstances, strongly compete with the RSFS. We shall
describe the latter effect in section \ref{neweffects}.

By inserting expression (\ref{expansion1}) into Eq. (\ref{substx1}),
and after integrating over the transverse coordinates of the fiber, we
can single out the following constant nonlinear coefficients:
\begin{eqnarray}
&&\Gamma^{jhpv}_{mkld}(\omega_{0})\equiv\frac{1}{16}\int
d\rr_{\perp}\chi^{(3)}_{xxxx}(\rrort)\cdot \nonumber\\
&&\cdot\frac{\left\{[\fff^{*(j)}_{m\omega_{0}}\cdot\fff^{(h)}_{k\omega_{0}}
]
[\fff^{(p)}_{l\omega_{0}}\cdot\fff^{*(v)}_{d\omega_{0}}]+[\fff^{*(j)}_{m\omega_{0}}\cdot\fff^{(p)}_{l\omega_{0}}
]
[\fff^{(h)}_{k\omega_{0}}\cdot\fff^{*(v)}_{d\omega_{0}}]+[\fff^{*(j)}_{m\omega_{0}}\cdot\fff^{*(v)}_{d\omega_{0}}
]
[\fff^{(h)}_{k\omega_{0}}\cdot\fff^{(p)}_{l\omega_{0}}]\right\}}{j!h!p!v!}\label{gammacapitalx2}
\end{eqnarray}

We shall analyze this complicated but very important object and its
large symmetries in the next section. $\Gamma^{jhpv}_{mkld}$ [in the
following, the upper indices $jhpv$ shall always indicate the order of
the derivatives in the Taylor expansion in (\ref{expansion1}), while
the lower indices $mkld$ shall always denote the polarization and mode
states] contains information about the nonlinearity experienced by the
different modes or polarizations in the fiber. Note that the structure
of  $\Gamma^{jhpv}_{mkld}$ closely resembles the one of the nonlinear
polarization $\PP_{\rm NL}$ inside Eq. (\ref{substx1}). In addition, all components
of $\Gamma^{jhpv}_{mkld}$ have the same physical dimensions. In particular, it is
convenient to use the usual units for nonlinear coefficients of $W^{-1}m^{-1}$ for the quantities $k_{0}\Gamma^{jhpv}_{mkld}$.

The final step in our derivation is to find an equation in the time
domain for the envelope of the electric field. In order to do this, we
decouple the fast oscillating phase by introducing the envelope
function $Q$, therefore using SVEA:
\eq{subst2}{A_{m\omega}\equiv
Q_{m,\Delta\omega}e^{i(\beta_{m\omega_{0}}-\beta_{m\omega})z}e^{-i\omega_{0}t}.}
In this paper we shall deal with perfectly circular photonic
nanowires, i.e. silica strands surrounded by air or vacuum. In order
to simplify our treatment we assume that the waveguide is operated in
a single-mode regime, so that the index $m=1,2$ only represents the
polarization states of the fundamental mode, which in perfectly
circular fibers are degenerate [$\beta_{1}(\omega)=\beta_{2}(\omega)$] and
orthogonal in the sense of the cross product, Eq. (\ref{orto2}).
This assumption can of course be relaxed without problems in our
formulation, but here we wish to focus on the novel properties of our
equation.
After some regrouping and a Fourier transformation over the
new independent variable $\Delta\omega$, we finally obtain our master equation:
\eq{finalx1}{i\de_{z}Q_{m}+\hat{D}_{m}(i\de_{t})Q_{m}+k_{0}\sum_{kld}\sum_{jhpv}\Gamma^{jhpv}_{mkld}\left[
G^{j}(t)\otimes \Phi^{hpv}_{kld}(t) \right]=0,} where
$k_{0}\equiv\omega_{0}/c$ is the vacuum wavenumber,
$\hat{D}_{m}(i\de_{t})\equiv\beta_{m}(\omega_{0}+i\de_{t})-\beta_{m}(\omega_{0})$
is the dispersion operator that encodes all the complexity of the
fiber GVD around the reference frequency \cite{kolesik}, and
\eq{finalx3}{G^{j}(\Delta\omega)\equiv\left(1+\frac{\Delta\omega}{\omega_{0}}\right)
\left(\frac{\Delta\omega}{\omega_{0}}\right)^{j}}
is a function that naturally contains the dynamics of the shock term
[through the term $(1+\Delta\omega/\omega_{0})$].
The convoluted nonlinear fields in Eq. (\ref{finalx1}) are given by:
\eq{finalx4}{\Phi^{hpv}_{kld}(t)\equiv\frac{\left[(i\de_{t})^{h}Q_{k}(t)\right]\left\{R(t)\otimes\left(\left[(i\de_{t})^{p}Q_{l}(t)\right]\left[(-i\de_{t})^{v}Q^{*}_{d}(t)\right]\right)\right\}}{\omega_{0}^{h+p+v}},}
that also depends on polarization state indices and derivative orders
in the expansion of Eq. (\ref{expansion1}).

Equation (\ref{finalx1}) is the central result of the present work. It
differs from other formulations in that it has been derived by making
no assumptions or approximations on the frequency dependence of the
field profiles of the linear modes of the waveguide. Higher-order derivatives of the mode profiles
embedded into Eq. (\ref{finalx1}) are responsible for the appearance
of additional nonlinear terms in the equations. These new terms obviously
become increasingly small with increasing orders of the derivatives,
since for every derivation there is a factor $1/\omega_{0}$. However,
as we demonstrate in section \ref{neweffects}, the new first-order
term has a strength comparable with or sometimes even larger than
the Raman and Kerr terms, and there are circumstances in which it must have observable consequences.

% reduction of master equation to known models
% analysis of Gamma and G
\section{Symmetries of the master equation and reduction to known models}
\label{symmetries}
The nonlinear coefficient $\Gamma^{jhpv}_{mkld}$ given in Eq.
(\ref{gammacapitalx2}) possesses very large symmetries that we wish to
analyze in this section, following similar arguments as in Ref. \cite{poletti}. Some of these symmetries are intrinsic, i.e.
valid for any fiber geometry, while some other are strongly
geometry-dependent. As we shall see, the symmetries of
$\Gamma^{jhpv}_{mkld}$ can be effectively used to reduce enormously
the complexity of Eq. (\ref{finalx1}). As a practical important
example, here we specialize the description of such symmetry relations
for the case of a silica nanowire with a perfectly circular core
embedded in air. In this case, we consider only the fundamental mode
of the fiber, which has two polarizations $m=1,2$ with the same
propagation constant [$\beta_{1}(\omega)=\beta_{2}(\omega)$]. In the
following we always assume that when index $m$ is fixed, then $n=3-m$
denotes the opposite polarization. As before, indices $j,h,p,v\geq 0$
are used to denote the orders of the derivatives in the definition
(\ref{gammacapitalx2}) of $\Gamma$, and they can take any positive
integer value.

Directly from the definition (\ref{gammacapitalx2}), one immediately
gets the relation $\Gamma^{jhpv}_{mkld}=\Gamma^{jphv}_{mlkd}$, so the
two lower an upper central indices can be simultaneously interchanged
without changing the value of the nonlinear coefficient. Other simple
relations (that can be directly verified by the reader) are the
following: $\Gamma^{jhpv}_{mklm}=\Gamma^{vphj}_{mlkm}$,
$\Gamma^{jhhh}_{mmnn}=\Gamma^{hjhh}_{mmnn}$,
$\Gamma^{jhhh}_{mmmm}=\Gamma^{hjhh}_{mmmm}$. All the above equalities
are always valid for any geometry of the waveguide.

In the special case of perfectly circular fiber, from direct
computation of Eq. (\ref{gammacapitalx2}) by using the explicit
formulas for the two (even and odd) degenerate polarizations of the
fundamental mode given in Ref. \cite{snyderlove}, one obtains the
useful equalities
$\Gamma^{jhpv}_{mkld}=\Gamma^{jhpv}_{(3-m)(3-k)(3-l)(3-d)}$,
$\Gamma^{jhpv}_{mkld}=\Gamma^{vhpj}_{dklm}$, and
$\Gamma^{jhhh}_{mnnm}=\Gamma^{hjhh}_{mnnm}$. Most importantly, many of
these coefficients vanish identically:
$\Gamma^{jhpv}_{mnmm}=\Gamma^{jhpv}_{mmnm}=\Gamma^{jhpv}_{mmmn}=\Gamma^{jhpv}_{nmmm}=\Gamma^{jhpv}_{nmnn}=\Gamma^{jhpv}_{nnmn}=\Gamma^{jhpv}_{nnnm}=\Gamma^{jhpv}_{mnnn}=0$.

We now demonstrate that our propagation equation (\ref{finalx1}) can
be reduced to the previously published model of Ref. \cite{monro1} by
using the above symmetries, and by taking into account only the lowest
order in the modal profile expansion Eq. (\ref{expansion1}). For this
purpose, let us write the zero-th order approximation ($j=h=p=v=0$) of
Eq. (\ref{finalx1}):
\eq{zeroth1}{i\de_{z}Q_{m}+\hat{D}(i\de_{t})Q_{m}+k_{0}\sum_{kld}\Gamma^{0000}_{mkld}\left[G^{0}\otimes\Phi^{000}_{kld}\right]=0.}
We now explicit the summations in Eq. (\ref{zeroth1}) for the
nonlinear functions $\Phi^{000}_{kld}$, obtaining:
\begin{eqnarray}
\lefteqn{i\de_{z}Q_{m}+\hat{D}(i\de_{t})Q_{m}+k_{0}\Gamma^{0000}_{mmmm}\left[G^{0}\otimes\left\{Q_{m}\left(R\otimes|Q_{m}|^{2}\right)\right\}\right]+k_{0}\Gamma^{0000}_{mnmm}
\left[G^{0}\otimes\left\{Q_{n}\left(R\otimes|Q_{m}|^{2}\right)\right\}\right]+}
\nonumber\\
& &k_{0}\Gamma^{0000}_{mmnm}\left[G^{0}\otimes\left\{Q_{m}\left(R\otimes
Q_{n}Q^{*}_{m}\right)\right\}\right]+k_{0}\Gamma^{0000}_{mmmn}\left[G^{0}\otimes\left\{Q_{m}\left(R\otimes
Q_{m}Q^{*}_{n}\right)\right\}\right]+\nonumber \\
& &k_{0}\Gamma^{0000}_{mnnn}\left[G^{0}\otimes\left\{Q_{n}\left(R\otimes
|Q_{n}|^{2}\right)\right\}\right]+
k_{0}\Gamma^{0000}_{mmnn}\left[G^{0}\otimes\left\{Q_{m}\left(R\otimes
|Q_{n}|^{2}\right)\right\}\right]+\nonumber \\
& &k_{0}\Gamma^{0000}_{mnmn}\left[G^{0}\otimes\left\{Q_{n}\left(R\otimes
Q_{m}Q^{*}_{n}\right)\right\}\right]+k_{0}\Gamma^{0000}_{mnnm}\left[G^{0}\otimes\left\{Q_{n}\left(R\otimes
Q_{n}Q^{*}_{m}\right)\right\}\right]\simeq 0. \label{example1}
\end{eqnarray}
According to the symmetries reported above, the fourth, fifth, sixth
and seventh term of Eq. (\ref{example1}) vanish identically. Moreover,
the eighth and ninth term have an identical nonlinear coefficient. A careful comparison
between all terms of Eq. (\ref{example1}) shows that, indeed, the
structure of this equation is identical to the one reported in Eq.
(45) of Ref. \cite{monro1}, in the limiting case of a single mode
regime, and without the Raman effect. However, our definition of the envelope $Q$ as given by Eq. (\ref{subst2})
automatically incorporates into the equations the frequency variations of the normalization constant (which is proportional to the effective mode area, see Ref. \cite{dudleyarea}),
something that has not been done in the approach of Ref. \cite{monro1}, where the constancy of $N_{m\omega}$ around $\omega_{0}$ was assumed.

% new terms and new effects derived from master equation
\section{New nonlinear effects in photonic nanowires} \label{neweffects}
\begin{figure}[htb]
\centering\includegraphics[width=7cm]{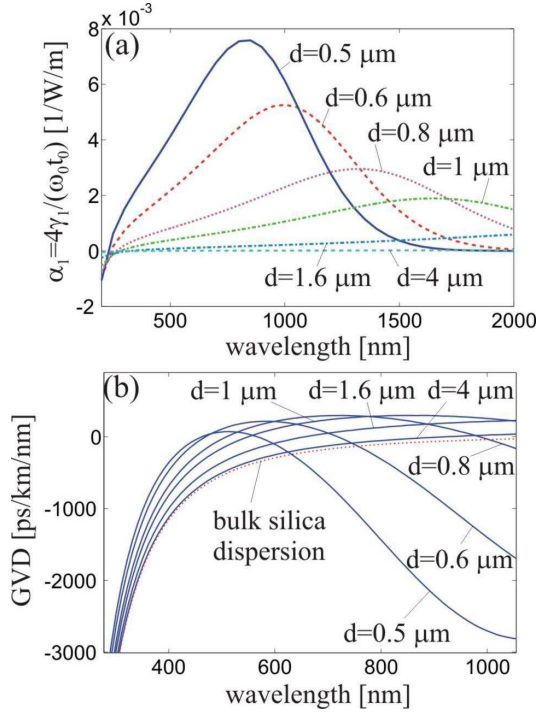}
\caption{\small (Color online) (a) Plots of calculated values of
$\alpha_{1}\equiv4\gamma_{1}/(\omega_{0}t_{0})$ in [m$^{-1}$W$^{-1}$]
describing the dispersion of the first-order nonlinear coefficient as
a function of wavelength, for silica strands in air of diameters
$d=0.5, 0.6, 0.8, 1, 1.6$ and $4$ $\mu$m in the fundamental mode of
one polarization ($m=1$). Pulse duration is
$t_{0}=100$ fs. For a given diameter $d$, the largest value of
$|\alpha_{1}|$ is located at approximately $\lambda\simeq Cd$, with $C\simeq 1.6$. (b)
Plots of calculated GVD for the same diameters as in (a). Red dashed
line indicates the GVD for bulk silica. A comparison between (a) and
(b) shows that, for silica strands, the maximum of $|\alpha_{1}|$ is
always located in the normal GVD region.} \label{fig3}
\end{figure}
The higher-order derivatives in the Taylor expansion performed in Eq.
(\ref{expansion1}) are expected to produce additional nonlinear terms,
previously unknown, that become increasingly important when the core
diameter is reduced and/or the refractive index contrast between cladding and core is increased. Here we attempt to
obtain a physical understanding of these new terms, at least at the
first-order in the Taylor expansion. In the following we just want
to isolate and understand as clearly as possible the new nonlinear
terms that come out of Eq. (\ref{finalx1}). In order to do this, we
just take one polarization state ($m=1$) for simplicity, we expand Eq.
(\ref{finalx1}) and collect all the zero-th and first-order terms
coming from the products between the $G$, $\Gamma$ and $\Phi$
functions, obtaining
\eq{effect1}{i\de_{z}Q+\hat{D}(i\de_{t})Q+k_{0}\Gamma^{0000}\Phi^{000}+\frac{i}{\omega_{0}}k_{0}\Gamma^{0000}\de_{t}\Phi^{000}+k_{0}\Gamma^{1000}\left[\frac{i}{\omega_{0}}\de_{t}\Phi^{000}+
\Phi^{100}+\Phi^{010}+\Phi^{001}\right]\simeq 0.} By using the
relation $(i/\omega_{0})\de_{t}\Phi^{000}=\Phi^{100}+\Phi^{010}-\Phi^{001}$
(that can be derived just by explicitly calculating the derivative of
$\Phi^{000}$), one can slightly simplify Eq. (\ref{effect1}) into
\eq{effect2}{i\de_{z}Q+\hat{D}(i\de_{t})Q+\gamma_{0}\left[1+\frac{i}{\omega_{0}}\left(1+2\frac{\gamma_{1}}{\gamma_{0}}\right)\de_{t}\right]\Phi^{000}+
2\gamma_{1}\Phi^{001}\simeq 0,} where we have defined the constants
$\gamma_{0}\equiv k_{0}\Gamma^{0000}$ and $\gamma_{1}\equiv
k_{0}\Gamma^{1000}$. A plot of the ratio
$r\equiv\Gamma^{1000}/\Gamma^{0000}=\gamma_{1}/\gamma_{0}$ is given in Fig.
\ref{fig2}(a) for silica strands of various diameters. Interestingly,
for a given diameter, $r$ is predominantly positive, and becomes negative only in regions of very short wavelengths, not easily accessible experimentally (typically below $200$ nm).
Also, $r$ is significantly larger than unity for wavelengths $\lambda>Cd$, where $C\simeq 1.6$ for perfectly circular silica strands in air.
In order to gain some insight into the physical meaning
of all nonlinear terms in Eq. (\ref{effect2}), we consider relatively
long pulses (of width $t_{0}\gg 100$ fs) for which the envelope
evolves slowly along the fiber. Under this assumption, one can
approximate the convolutions with nonlinear combinations of the field
$Q$ and its derivatives \cite{agrawalbook}. In particular, one has
\begin{eqnarray}
\Phi^{000}&=&Q(z,t)\int_{-\infty}^{+\infty}R(t')|Q(z,t-t')|^{2}dt'\simeq
Q(z,t)\int_{-\infty}^{+\infty}R(t')\left\{|Q(z,t)|^{2}-t'\de_{t}|Q(z,t)|^{2}\right\}dt'= \nonumber \\
& &=|Q|^{2}Q-T_{R}Q\de_{t}|Q|^{2}, \label{convol1} \\
\Phi^{001}&=&-\frac{i}{\omega_{0}}Q(z,t)\int_{-\infty}^{+\infty}R(t-t')Q(z,t')\de_{t'}Q^{*}(z,t')dt'= \nonumber \\
&-&\frac{i}{\omega_{0}}Q(z,t)\int_{-\infty}^{+\infty}R(t')Q(z,t-t')\left[\frac{\de
Q^{*}(z,\tau)}{\de\tau}\right]_{\tau=t-t'}dt'\simeq \nonumber\\
&\simeq& -\frac{i}{\omega_{0}}Q(z,t)\int_{-\infty}^{+\infty}R(t')\left[Q(z,t)-t'\de_{t}Q(t)\right]\left[\frac{\de
Q^{*}(t)}{\de t}-t'\frac{\de^{2}Q^{*}(t)}{\de t^{2}}\right]dt'\simeq \nonumber\\
&\simeq&-\frac{i}{\omega_{0}}\left[Q^{2}\de_{t}Q^{*}-T_{R}Q^{2}\de_{t}^{2}Q^{*}-T_{R}Q|\de_{t}Q|^{2}+T_{R}'Q(\de_{t}Q)(\de_{t}^{2}Q^{*})\right], \label{convol2}
\end{eqnarray}
where $T_{R}\equiv\int_{-\infty}^{+\infty}t'R(t')dt'$ is the first
moment of the Raman nonlinear response function, and
$T_{R}'\equiv\int_{-\infty}^{+\infty}t'^{2}R(t')dt'$ is the second
moment. The above simple Lorentzian model for the Raman response
function gives $T_{R}\simeq 1.5$ fs and $T_{R}'\simeq -23$ fs$^{2}$
for silica fibers.
Substituting Eqs. (\ref{convol1}-\ref{convol2}) into Eq.
(\ref{effect2}), and neglecting the second order derivatives and all
terms of order of $\omega_{0}^{-2}$ or $T_{R}/\omega_{0}$, after some algebra we obtain
our final expression (in the first-order of the perturbation theory)
for the approximate evolution equation for long pulses:
\eq{approx1}{i\de_{z}Q+\hat{D}(i\de_{t})Q+\gamma_{0}\left[\left(\underbrace{1+\frac{i}{\omega_{0}}\de_{t}}_{T0}\right)\underbrace{|Q|^{2}Q}_{T1}-\underbrace{T_{R}Q\de_{t}|Q|^{2}}_{T2}\right]+\underbrace{\frac{4i\gamma_{1}}{\omega_{0}}|Q|^{2}\de_{t}Q}_{T3}\simeq
0.}
In Eq. (\ref{approx1}), the term indicated by T0 represents the
well-known shock operator \cite{agrawalbook}.
Terms T1 and T2 in Eq. (\ref{approx1}) correspond to the usual and
easily recognizable Kerr effect and RSFS
respectively \cite{agrawalbook,moment}. The additional term T3 is
connected to a {\em nonlinear} change of the group velocity,
proportional to the field intensity $|Q|^{2}$. Such nonlinear term arises quite unexpectedly, and it is
strictly related to the way in which the mode profile changes with
frequency. Slightly different frequencies have different linear mode
profiles, which in turn will have slightly different group velocities.
This change becomes nonlinear, because the time derivative inside $\phi^{001}$
in Eq. (\ref{finalx4}) is embedded into the full nonlinear convolution of Eq. (\ref{finalx4}), thus giving rise
to term T3 in Eq. (\ref{approx1}). It is now easy to see that ratio $r$ of Fig. \ref{fig2}(a) quantifies the
relative importance between the new nonlinearity T3 and the Kerr nonlinearity T1 in Eq. (\ref{approx1}).
Note that $r$ can be as large as $6$ in silica strands inside the frequency window accessible experimentally.

The relative importance between the T3 and T2 terms is regulated by a
dimensionless parameter $\alpha_{S}\equiv
4\gamma_{1}/(\gamma_{0}\omega_{0}T_{R})$, which is shown in Fig.
\ref{fig2}(b) as a function of wavelength for silica strands of
various diameters. Again, this function is positive above $200$ nm.
Note that $\alpha_{S}$ can be as large as $15$ inside the frequency window accessible experimentally.

The presence of the extra first-order term T3 has not been previously
reported in the context of unidirectional pulse propagation equations,
and it constitutes an important result of the present work. It can be demonstrated \cite{future} that this term
may lead to either a suppression or an enhancement of the RSFS,
depending on whether the coefficient $\alpha_{S}$ is respectively
positive or negative. In addition, the discussions above demonstrate that there are broad regions of wavelengths in which T3 can be several times
larger than both the Kerr and the Raman terms.
Fig. \ref{fig3}(a) shows the calculated value of
the new first-order nonlinear coefficient
$\alpha_{1}\equiv4\gamma_{1}/(\omega_{0}t_{0})$ as a function of
wavelength, for silica strands in air of various core sizes and for
$t_{0}=100$ fs. It is
interesting to see that the optimal wavelength for which
$|\alpha_{1}|$ is largest approximately corresponds to $\lambda\simeq
Cd$, again with $C\simeq 1.6$. Moreover, it can be seen in Fig. \ref{fig3}(a) that such a linear law
seems to be a universal behavior, once that geometry and core and cladding materials are specified.
Fig. \ref{fig3}(b) shows the calculated GVD of silica strands in air
of various core sizes. By comparing Fig. \ref{fig3}(b) with Fig.
\ref{fig3}(a) one can see that the condition $\lambda\simeq Cd$ is always
satisfied in correspondence of normal GVD for this kind of circular geometry.
In our extensive numerical simulations we have seen that the
new first-order nonlinear terms make a substantial difference in the
pulse propagation only when the input pulse is launched in the
anomalous dispersion, i.e. in presence of soliton formation, where it can enter in competition
with the RSFS of solitons, while in
the normal dispersion regime the contribution of the term proportional
to $\phi^{001}$ on the dynamics is largely negligible. For this reason, it
seems that silica circular nanowires are not optimal for evidencing
the effects of the new nonlinear term T3 regulated by nonlinear
coefficient $\alpha_{1}$. Possibly this is why such new effects have never previously been noticed experimentally, although techniques to fabricate
photonic nanowires have already been available since a few years \cite{foster1}.
A detailed study of the optimization of the
new nonlinear terms in sub-wavelength-core PCFs and high-refractive
index glasses will be carried out in a separate publication
\cite{future}.

As a final note, we would like to stress again that the expansion of
Eq. (\ref{convol2}) is valid only for relatively long pulses. For
ultra-short pulses, such as $t_{0}<50$ fs, all the physical effects
and terms that are clearly visible in the formulation of Eq.
(\ref{convol2}) will be naturally embedded into the full convolution
contained in $\Phi^{001}$, see Eq. (\ref{effect2}).

% numerical analysis; discussion and comparison with other models
\section{Numerical analysis and comparison with previous models}
\label{numericalanalysis}
In this section we show results of the extensive numerical simulations
that was performed by using Eq. (\ref{finalx1}). We shall compare the
output spectra obtained by pulse propagation in different dispersive
regimes, and for the situation in which Eq. (\ref{finalx1}) is
truncated at the zero-th order, as opposed as when instead all terms
of first-order are included.

As our representative case, we take a silica strand of core diameter
$d=0.6$ $\mu$m in air, the dispersion of which is displayed in Fig.
\ref{fig3}(b). For this diameter, the two zero GVD points are approximately located
at $\lambda_{1}=0.468$ $\mu$m and $\lambda_{2}=0.718$
$\mu$m. The input pulse duration is always taken to have a temporal
width equal to $t_{0}=100$ fs, with the shape of a hyperbolic secant,
and with an input power $P$ such that the 'number of solitons'
parameter $N\equiv[\gamma_{0}Pt_{0}^{2}/|\beta_{2}(\lambda_{0})|]^{1/2}$
is always equal to $N=7$, where $\beta_{2}(\lambda_{0})$ is the GVD
coefficient calculated at the pump wavelength $\lambda_{0}$.
Propagation distances are given in dimensionless units: $\xi\equiv
z/L_{D2}(\lambda_{0})$, where $L_{D2}(\lambda_{0})\equiv
t_{0}^{2}/|\beta_{2}(\lambda_{0})|$ is the second-order dispersion
length at the pump wavelength $\lambda_{0}$ \cite{agrawalbook}. Also,
for simplicity we assume for the moment that only one polarization
state is excited ($m=1$), and we neglect the dynamics of the other
polarization state. In the final part of this section we shall also
show some result obtained when taking into account the full
polarization structure of Eq. (\ref{finalx1}).

\begin{figure}[htb]
\centering\includegraphics[width=12cm]{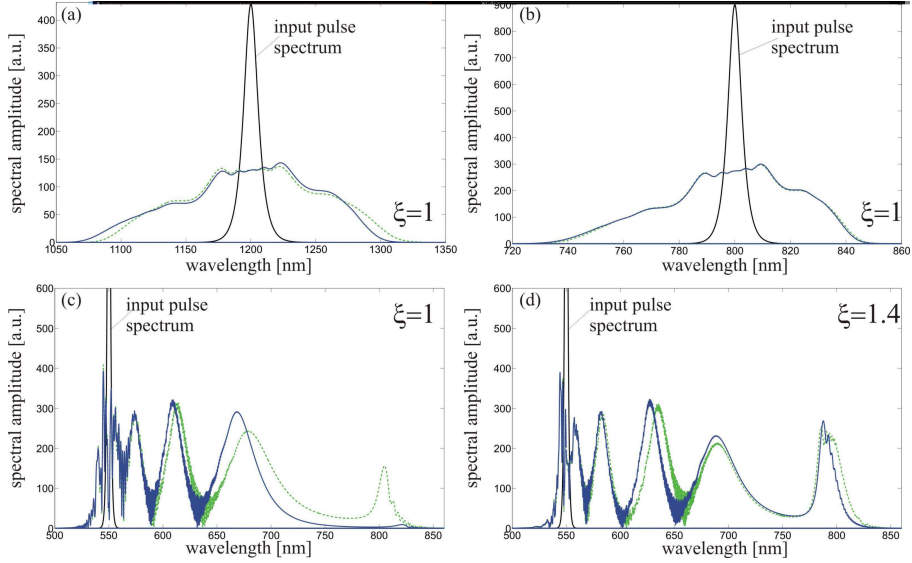}
\caption{\small (Color online) Simulation of light propagation in a
photonic nanowire of diameter $d=0.6$ $\mu$m, for only one
polarization state ($m=1$). Input pulse duration is $t_{0}=100$
fs, and input pulse power is always $N=7$. (a) Output spectra of the
simulated Eq. (\ref{finalx1}) in normal dispersion regime after a
propagation length of $\xi=1$, by using different truncations of the Taylor series
contained into Eq. (\ref{approx1}).
Input pulse is centered at $\lambda_{0}=1.2$ $\mu$m (black solid
line). Blue solid line indicates the solution of Eq.
(\ref{finalx1}) when truncating the sum up to the first-order in the
Taylor expansion. Green dashed line gives the same as the blue solid
line, but when the sum in Eq. (\ref{finalx1}) is truncated at the
zero-th order of expansion, i.e. when neglecting the influence of the
new first-order nonlinear terms described in section \ref{neweffects}.
(b) Same as in (a), but for a pump wavelength
$\lambda_{0}\simeq 0.8$ $\mu${m}. (c)
Same as (a), but pumping at a wavelength $\lambda_{0}=0.55$ $\mu$m in
the anomalous dispersion, for $\xi=1$. (d) Same as (c),
but for $\xi=1.4$.} \label{fig4}
\end{figure}

Fig. \ref{fig4}(a) shows the output spectrum of an input pulse (displayed
in black solid line) propagating in the normal dispersion regime of
the nanowire, when pumping at $\lambda_{0}=1.2$ $\mu$m, and after a
dimensionless propagation distance of $\xi=1$ ($L_{D2}\simeq 6$ mm). At this wavelength we have $r=1.351$ and $\alpha_{S}=2.295$, see Fig. \ref{fig2}(a,b). The blue solid line in
Fig. \ref{fig4}(a) shows the result of the simulation of equation
(\ref{finalx1}), when considering the expansion of the Taylor series
of the mode profiles up to the first order. The green dashed line in Fig. \ref{fig4}(a) shows
the results of the simulation of Eq. (\ref{finalx1}), when considering
only the zero-th order in the expansion. It is possible to observe that in the case of
normal dispersion there is indeed only a small difference between the
green dashed line and the blue solid line, which means that the new
first-order nonlinear terms contained in Eq. (\ref{finalx1}) give only
a relatively small contribution in this regime, as anticipated before.

When launching the input pulse at $\lambda_{0}=0.8$ $\mu$m ($L_{D2}\simeq 10$ cm), still in
the normal dispersion regime but closer to one of the zero GVD points,
all three curves nearly coincide, see Fig. \ref{fig4}(b). In fact, the new nonlinear coefficient is smaller than in the
previous case, $r=0.3973$ and $\alpha_{S}=0.45$. Thus we can conclude that in absence of soliton dynamics, the two models give qualitatively and
quantitatively very similar results.

A completely different scenario is observed in presence of solitons, see Fig.
\ref{fig4}(c). This figure shows the output spectrum of a pulse
launched in the anomalous dispersion regime at $\lambda_{0}=0.55$
$\mu$m, for a propagation of $\xi=1$ ($L_{D2}\simeq 30$ cm). At this wavelength, $r=0.1537$ and $\alpha_{S}=0.1196$.
Fig. \ref{fig4}(c) shows that in the initial stages of the propagation, the stabilization
of RSFS by means of resonant radiation (at around $800$ nm in the
figure) emitted near the zero GVD point \cite{scienceskryabin}, occurs
earlier when the equations do not take into account the new first-order nonlinear term
T3 in Eq. (\ref{approx1}), as the strongest soliton red-shifts at a faster rate in this latter (less precise) model.
It was mentioned in section \ref{neweffects} that the novel nonlinear terms
associated to the convolution $\phi^{001}$ in Eq. (\ref{finalx1}) can give rise either to a suppression or
an enhancement of the RSFS of solitons, depending on whether the sign of coefficient
$\alpha_{S}$ is positive or negative, respectively. This is due to a peculiar interplay between the Raman effect
and the nonlinear change of group velocity induced by the new terms in the solitonic regime \cite{future}. Despite the smallness of coefficients $r$ and $\alpha_{S}$ at this wavelength, a quite strong suppression scenario is
already visible in Fig. \ref{fig4}(c). In fact, one can see that
in the full model that takes into account the additional first-order nonlinear terms (blue solid line) solitons are pushed towards longer wavelengths at a much slower rate than in the
model without first-order nonlinear effects (green dashed line). If the pump
wavelength is located in a frequency region where $\alpha_{S}$ is negative, exactly the opposite scenario would be observed, and the blue solid line
would be shifted towards longer wavelengths with respect to the green dashed line, which means that the RSFS rate is increased. However, such a region of positive
$\alpha_{S}$ is located in the UV in circular silica strands [see Fig. \ref{fig2}(b)] and it is not reachable experimentally. Thus, for the geometry considered in this paper,
the suppression of RSFS is the only physically meaningful scenario.
Due to the relative smallness of $|\alpha_{S}|$ inside the anomalous dispersion
regime of the circular silica waveguides under consideration, the corrections to RSFS induced by the new nonlinear terms are also not as large as they could be. However, as we shall show in detail in a future publication \cite{future}, by using properly designed PCFs and high-refractive index materials, one is able to bring the wavelength for which $|\alpha_{1}|$ has a maximum [see Fig. \ref{fig3}(a)] {\em inside} the anomalous dispersion regime. In such a case, the new nonlinear terms will completely change the dynamics of solitons, leading to qualitatively new regimes of nonlinear propagation in which
the RSFS of solitons can be completely stopped for many dispersion lengths, without the use of stabilization techniques based on solitonic resonant dispersive wave emission \cite{scienceskryabin}.

In the same conditions as in Fig. \ref{fig4}(c), but for a slightly
longer propagation distance $\xi=1.4$ [see Fig. \ref{fig4}(d)], the solid blue and the dashed green lines
become qualitatively very similar, apart from the visible
differences in the dynamics of generation of resonant radiation,
around $800$ nm in Fig. \ref{fig4}(d), the bandwidth of which is
somewhat reduced in the model truncated at the first-order (blue solid line) than in the model truncated at the
zeroth-order (green dashed line).

\begin{figure}[htb]
\centering\includegraphics[width=10cm]{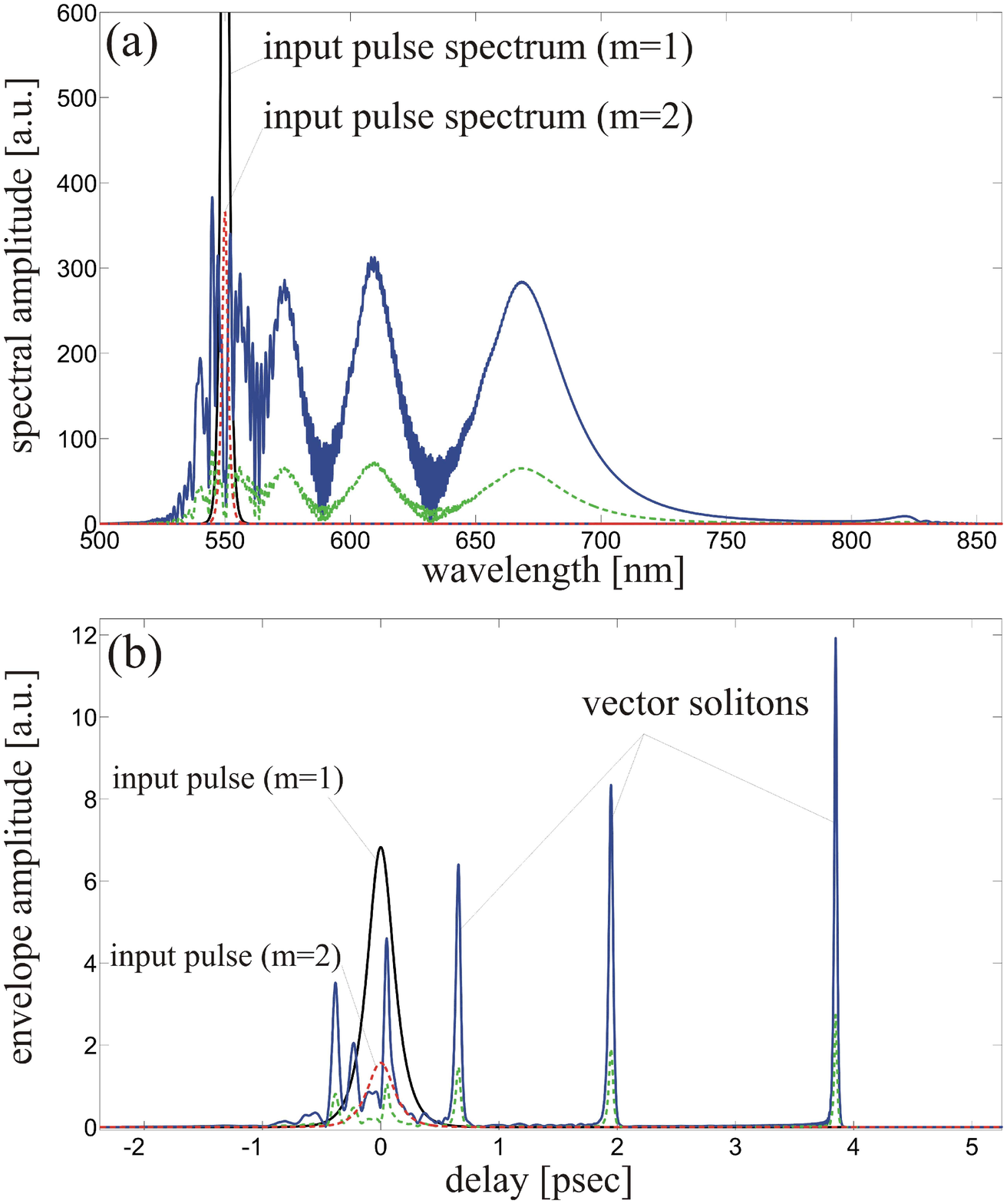}
\caption{\small (Color online) (a) Final spectrum of a linearly
polarized input pulse ($95\%$ on the $m=1$ axis and $5\%$ along the
$m=2$ axis), as obtained by simulating the full Eq. (\ref{finalx1}). All
other parameters for the pulse and the fiber are the same as in Fig.
\ref{fig4}(c). Black solid and red dashed lines are respectively the
polarization components of the input pulse along $m=1$ and $m=2$
respectively. Blue solid line is output pulse along $m=1$, while green dashed line is along $m=2$. (b) Time domain picture of the propagation
shown in (a). The formation of several vector solitons of different
amplitudes is observed, each of them subjected to RSFS according to
their individual intensities.} \label{fig5}
\end{figure}

In all the above calculations we have simulated only one of the two
polarization states, in order to clearly display the differences
between several models and approximations. However, Eq.
(\ref{finalx1}) has a vector nature and actually represents a pair of
coupled equations for the two polarization fields $Q_{1}$ and $Q_{2}$.
In Fig. \ref{fig5}(a) the final spectrum of a linearly polarized input
pulse ($95\%$ on the $m=1$ axis and $5\%$ along the $m=2$ axis) is
shown, simulated by using the full Eq. (\ref{finalx1}). All other
parameters for the pulse and the fiber are the same as in Fig.
\ref{fig4}(c). Fig. \ref{fig5}(b) shows the time domain picture which
corresponds to Fig \ref{fig5}(a). As expected, the formation of several vector
solitons of different amplitudes is observed, each of them subject
to RSFS according to their individual intensities.

% conclusions and future work
\section{Conclusions and future work}
\label{conclusions}
In conclusion, we have formulated a new set of vector unidirectional
evolution equations [Eq. (\ref{finalx1})] for silica photonic
nanowires, i.e. for waveguides with a core smaller that the wavelength
of light, and with a large step in the refractive indices of the core
and the cladding, for which the $z$-component of the electric field is
not negligible.
Our model differs from previous formulations in that we
simultaneously take into account the vector nature of the EM field and
the variations of the linear mode profiles with frequency. Such
variations generate new higher-order nonlinear terms,
given by the Taylor expansion in Eq. (\ref{finalx1}). An analysis of
the long-pulse limit of Eq. (\ref{finalx1}) has been carried out,
which makes the physical understanding of the new first-order
nonlinear terms implicitly contained in Eq. (\ref{finalx1}) much more
transparent. This analysis led to Eq. (\ref{approx1}), in which a
nonlinear group velocity term appears, the strength of which is
regulated by coefficient $\alpha_{1}$, which for silica strands in air
of given diameter $d$ takes appreciable values in the normal
dispersion for wavelengths $\lambda>Cd$, with $C\simeq 1.6$. Numerical simulations
of pulse propagation in silica strands were provided. The difference
between the output spectra of our model and previous formulations is
significant, especially when the propagation occurs in the anomalous
dispersion regime, where soliton dynamics is mostly affected by the new nonlinearities.

Future works include the use of ultrasmall solid-core PCFs and interstitial
features in the cladding of hollow-core PCFs, as well as high-refractive index nanowires, for the optimization of
the novel nonlinear terms. This optimization, which makes use of the
great flexibility in the dispersion tailoring of the above fibers,
will bring the wavelength range at which the new nonlinearities are
most effective (i.e. where $|\alpha_{1}|$ is around its maximum)
inside the anomalous dispersion regime of the fiber. This will create
a truly qualitatively new and unexplored nonlinear behavior in such
waveguides.

This work is supported by the German Max Planck Society for the
Advancement of Science (MPG). The authors would like to acknowledge Prof. P.
St. J. Russell for stimulating discussions, and S. Stark for
helping with the numerical codes.

%%%%%%%%%%%%%%%%%%%%%%% References %%%%%%%%%%%%%%%%%%%%%%%%%


\begin{thebibliography}{99}

\bibitem{agrawalbook} G. P. Agrawal, {\em Nonlinear Fiber Optics}, 4th ed. (Academic Press, San Diego, 2007).

\bibitem{allentaflovebook} A. Taflove and S. C. Hagness, {\em Computational Electrodynamics}, 3rd ed. (Artech House, London, 2005).

\bibitem{hasegawakodama} A. Hasegawa and Y. Kodama, ``Nonlinear pulse propagation in a monomode dielectric guide,'' IEEE J. Quantum Electron. {\bf 23,} 510--524 (1987).

\bibitem{modulationalinstability} F. Biancalana, D. V. Skryabin, and P. St. J. Russell, ``Four-wave mixing instabilities in photonic crystal and tapered fibers,'' Phys. Rev. E {\bf 68,} 046603 (2003).

\bibitem{gordon} J. P. Gordon, ``Theory of the soliton self-frequency shift,'' Opt. Lett. {\bf 11,} 662--664 (1986).

\bibitem{gaeta} A. L. Gaeta, ``Nonlinear propagation and continuum generation in microstructured optical fibers,'' Opt. Lett. {\bf 27,} 924--926 (2002).

\bibitem{taperedfibers} K. Shi, F. G. Omenetto, and Z. Liu, ``Supercontinuum generation in an imaging fiber taper,'' Opt. Express {\bf 14,} 12359--12364 (2006); C. M. B. Cordeiro, W. J. Wadsworth, T. A. Birks, P. St. J. Russell, ``Engineering the dispersion of tapered fibers for supercontinuum generation with a 1064 nm pump laser,'' Opt. Lett. {\bf 30,} 1980--1982 (2005).

\bibitem{russell} P. St. J. Russell, ``Photonic crystal fibers,'' Science {\bf 299,} 358--362 (2003).

\bibitem{dudley} J. M. Dudley, G. Genty and S. Coen, ``Supercontinuum generation in photonic crystal fibers,'' Rev. Mod. Phys. {\bf 78,} 1135--1184 (2006).

\bibitem{akhmediev} N. Akhmediev and M. Karlsson, ``Cherenkov radiation emitted by solitons in optical fibers,'' Phys. Rev. A {\bf 51,} 2602--2607 (1995).

\bibitem{scienceskryabin} D. V. Skryabin, F. Luan, J. C. Knight, and P. St. J. Russell, ``Soliton self-frequency shift cancellation in photonic crystal fibers,'' Science  {\bf 301,} 1705--1707 (2003).

\bibitem{biancalana} F. Biancalana, D. V. Skryabin, and A. V. Yulin, ``Theory of the soliton self-frequency shift compensation by the resonant radiation in photonic crystal fibers,'' Phys. Rev. E {\bf 70,} 016615 (2004).

\bibitem{brabec} T. Brabec and F. Krausz, ``Nonlinear optical pulse propagation in the single-cycle regime,''  Phys. Rev. Lett. {\bf 78,} 3282--3285 (1997).

\bibitem{mercedes}  T. G. Euser, J. S. Y. Chen, M. Scharrer, P. St. J. Russell, N. J. Farrer and P. J. Sadler, ``Quantitative broadband chemical sensing in air-suspended solid-core fibers,'' J. Appl. Phys. {\bf 103}, 103108 (2008).

\bibitem{benabid} F. Benabid, F. Biancalana, P. S. Light, F. Couny, A. Luiten, P. J. Roberts, J. Peng, A. V. Sokolov, ``Fourth-order dispersion mediated solitonic radiations in HC-PCF cladding,'' Opt. Lett. {\bf 33,} 2680--2682 (2008).

\bibitem{monro1} S. Afshar V. and T. M. Monro, ``A full vectorial model for pulse propagation in emerging waveguides with sub-wavelength structures part I: Kerr nonlinearity,'' Opt. Express {\bf 17,} 2298--2318 (2009).

\bibitem{monro2} M. D. Turner, T. M. Monro and S. Afshar V., ``A full vectorial model for pulse propagation in emerging waveguides with sub-wavelength structures part II: Stimulated Raman Scattering,'' Opt. Express {\bf 17,} 11565--11581 (2009).

\bibitem{foster1} M. A. Foster, A. C. Turner, M. Lipson, and A. L. Gaeta, ``Nonlinear optics in photonic nanowires,'' Opt. Express {\bf 16,} 1300--1320 (2008).

\bibitem{foster2} M. A. Foster, K. D. Moll, and A. L. Gaeta, ``Optimal waveguide dimensions for nonlinear interactions,'' Opt. Express {\bf 12,} 2880--2887 (2004).

\bibitem{zheltikov} A. Zheltikov, ``Gaussian-mode analysis of waveguide-enhanced Kerr-type nonlinearity of optical fibers and photonic wires,'' J. Opt. Soc. Am. B {\bf 22,} 1100--1104 (2005).

\bibitem{conferencemonro} S. Afshar V., W. Zhang and T. M. Monro, ``Experimental confirmation of a generalized definition of the effective nonlinear coefficient in emerging waveguides with sub-wavelength structures,'' CThBB6, CLEO Conference, Baltimore, USA (2009).


\bibitem{mamyshev} P. V. Mamyshev and S. V. Chernikov, ``Ultrashort-pulse propagation in optical fibers,'' Opt. Lett. {\bf 15,} 1076--1078 (1990).

\bibitem{kolesik} M. Kolesik and J. V. Moloney, ``Nonlinear optical pulse propagation simulation: From Maxwell's to unidirectional equations,'' Phys. Rev. E {\bf 70,} 036604 (2004); M. Kolesik, E. M. Wright, and J. V. Moloney, ``Simulation of femtosecond pulse propagation in sub-micron diameter tapered fibers,'' Appl. Phys. B {\bf 79,} 293--300 (2004).

\bibitem{laegsgaard} J. L{\ae}gsgaard, ``Mode profile dispersion in the generalised nonlinear Schršdinger equation,'' Opt. Express {\bf 15,} 16110--16123 (2007).

%\bibitem{blowwood} K. J. Blow and D. Wood, ``Theoretical description of transient stimulated Raman scattering
%in optical fibers,'' IEEE J. Quantum Electron. {\bf 25}, 2665 (1989).

%\bibitem{karasawa} N. Karasawa, S. Nakamura, N. Nakagawa, M. Shibata, R. Morita, H. Shigekawa and
%M. Yamashita, ``Comparison between theory and experiment of nonlinear propagation for a-few-cycle and
%ultrabroadband optical pulses in a fused-silica Fiber,'' IEEE J. Quantum Electron. {\bf 37}, 398 (2001).

\bibitem{jacksonbook} J. D. Jackson, {\em Classical Electrodynamics} (Wiley \& Sons, New York, 1998).

\bibitem{snyderlove} A. W. Snyder and J. Love, {\em Optical Waveguide Theory} (Kluwer, Boston, 1983).

\bibitem{dudleyarea} B. Kibler, J. M. Dudley, S. Coen, ``Supercontinuum generation and nonlinear pulse propagation in photonic crystal fiber: influence of the frequency-dependent effective mode area,'' Appl. Phys. B {\bf 81} 337--342 (2005).

\bibitem{boydbook} R. W. Boyd, {\em Nonlinear Optics}, 3rd ed. (Academic Press, San Diego, 2008).

\bibitem{hellwarth} R. W. Hellwarth, ``Third-order optical susceptibilities of liquids ans solids,'' Prog. Quantum Electron. {\bf 5,} 1--68 (1977).

\bibitem{poletti} F. Poletti and P. Horak, ``Description of ultrashort pulse propagation in multimode optical fibers,'' J. Opt. Soc. Am. B 25, 1645-1654 (2008).


%moment method
\bibitem{moment} J. Santhanam and G. P. Agrawal, ``Raman-induced spectral shifts in optical fibers: general theory based on the moment method,'' Opt. Commun. {\bf 222,} 413--420 (2003).

\bibitem{future} F. Biancalana and Tr. X. Tran, in preparation.


\end{thebibliography}
\end{document}